\documentclass[onecolumn, 12pt, draftclsnofoot]{IEEEtran}
\usepackage{amsfonts}
\usepackage{algorithmic}
\usepackage{algorithm}
\usepackage{amsmath}
\usepackage{amsthm}
\usepackage{array}
\usepackage{graphicx}
\usepackage{color}
\usepackage{multicol}
\usepackage{amssymb}
\usepackage{subfigure}
\usepackage{bm}
\usepackage{latexsym}
\usepackage{stfloats}
\usepackage{cite}
\usepackage{url}
\usepackage{soul}

\begin{document}

\title{Machine Learning for Vehicular Networks}
\author{Hao Ye, Le Liang, and Geoffrey Ye Li, Georgia Institute of Technology\\
JoonBeom Kim, Lu Lu, and May Wu, Intel Corporation

\thanks{H. Ye, L. Liang and G. Y. Li are with the School of Electrical and Computer Engineering, Georgia Institute of Technology, Atlanta, GA 30339 USA (e-mail: \{yehao,lliang\}@gatech.edu; liye@ece.gatech.edu).}

\thanks{J. Kim, L. Lu, and M. Wu are with Intel Corporation, Hillsboro, OR 97124 USA.}
}

\maketitle

\begin{abstract}
The emerging vehicular networks are expected to make everyday vehicular operation safer, greener, and more efficient, and pave the path to autonomous driving in the advent of the fifth generation (5G) cellular system.
Machine learning, as a major branch of artificial intelligence, has been recently applied to wireless networks to provide a data-driven approach to solve traditionally challenging problems.
In this article, we review recent advances in applying machine learning in vehicular networks and attempt to bring more attention to this emerging area.
After a brief overview of the major concepts of machine learning, we present some application examples of machine learning in solving problems arising in vehicular networks.
We finally discuss and highlight several open issues that warrant further research.
\end{abstract}

\section{Introduction}
With vehicles becoming more aware of their environments and evolving towards full autonomy, it calls for a new level of connectivity among them, leading to the  concept of connected vehicles.
As such, the emerging vehicular network has been regarded as an important component of the development of the intelligent transportation system (ITS) and smart cities.
It is expected to enable a whole new set of applications, ranging from road safety improvement to traffic efficiency optimization, from autonomous driving to ubiquitous Internet access on vehicles \cite{Liang2017vehicular,Peng2017vehicular}.
This new generation of networks will ultimately have a profound impact on society and the everyday lives of millions of people around the world.

Despite its significant potential to transform everyday vehicular experience, the vehicular network also brings unprecedented challenges unseen in traditional wireless communications systems due to the strict and diverse quality of service (QoS) requirements as well as the inherent dynamics in vehicular environments, such as fast varying wireless propagation channels and ever changing network topology.
To tackle such challenges, various communication standards, e.g., dedicated short range communications (DSRC) in the United States and the ITS-G5 in Europe, both based on the IEEE 802.11p standard, have been developed across the globe \cite{Liang2017vehicular}. Recently, the 3rd Generation Partnership Project (3GPP) has also started projects towards supporting vehicle-to-everything (V2X) services in long term evolution (LTE) networks and the future fifth generation (5G) cellular system \cite{Liang2017vehicular,Peng2017vehicular}.

\begin{figure}[!t]
\centering
\includegraphics[width=0.7\linewidth]{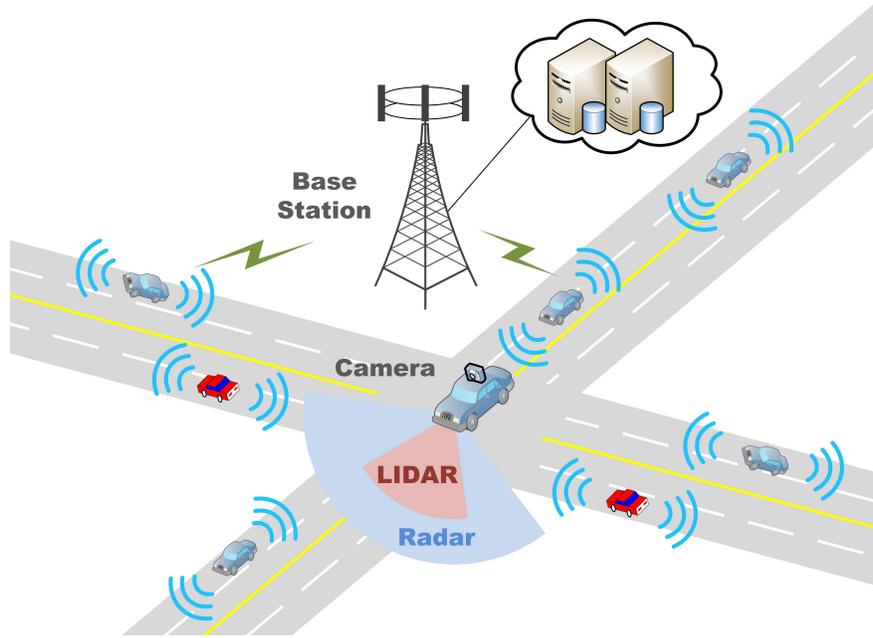}
\caption{An illustrative structure of vehicular networks.}\label{fig:sysFig}
\end{figure}

Meanwhile, with high performance computing and storage facilities and various advanced onboard sensors equipped, such as light detection and ranging (LIDAR), radars, and cameras, vehicles will be more than just a simple means of transportation. They are generating, collecting, storing, processing, and transmitting massive amounts of data designed to make driving safer and more convenient, as illustrated in Fig.~\ref{fig:sysFig}. These rich sources of data will necessarily provide new dimensions and abundant opportunities to explore for the design of reliable and efficient vehicular networks.
However, traditional communication strategies are not designed to handle and exploit such rich information.
Machine learning, as a major branch of artificial intelligence (AI), builds intelligent systems to operate in complicated environments and has found many successful applications in computer vision, natural language processing, and robotics \cite{MLIntro,DLIntro}.
It develops efficient methods to analyze a huge amount of data by finding patterns and underlying structures, {which can be beneficial to supporting the future smart radio terminals \cite{Han}}.
Moreover, machine learning represents an effective data-driven approach, making it more robust to handle heterogeneous data since no explicit assumptions are made on the data distribution.
As such, machine learning provides a versatile set of tools to exploit and mine multiple sources of data generated in vehicular networks.
This will help the system make more informed and data-driven decisions and alleviate communications challenges as well as providing unconventional services, such as location-based services, real-time traffic flow prediction and control, autonomous driving, etc.
However, how to adapt and exploit such tools to serve the purpose of vehicular networks remains a challenge and represents a promising research direction.
The objective of this article is to bring more attention to this emerging field since the research on applying machine learning in vehicular networks is still in its infancy.

In this article, we begin with reviewing the basic idea and algorithms of machine learning, and then introduce some preliminary examples of applying machine learning to address data-driven decision making and wireless resource management problems in vehicular networks.
We conclude with a discussion of some open challenges in this area for further research.

\section{Machine Learning Tools}
Machine learning methods can be roughly divided into three major categories: supervised learning, unsupervised learning, and reinforcement learning.
Table~\ref{tab:ALL} illustrates the family tree of machine learning with applications in wireless networks.
Other learning schemes, such as semi-supervised learning, online learning, and transfer learning, can be viewed as variants of these three basic types.
In general, machine learning involves two stages, i.e., training and testing.
In the training stage, a model is learned based on training data while in the testing stage, the trained model is applied to produce the prediction.

\begin{table}[!t]
\centering
\caption{Overview of machine learning algorithms and applications in wireless networks. }\label{tab:ALL}
\includegraphics[width=0.8\linewidth]{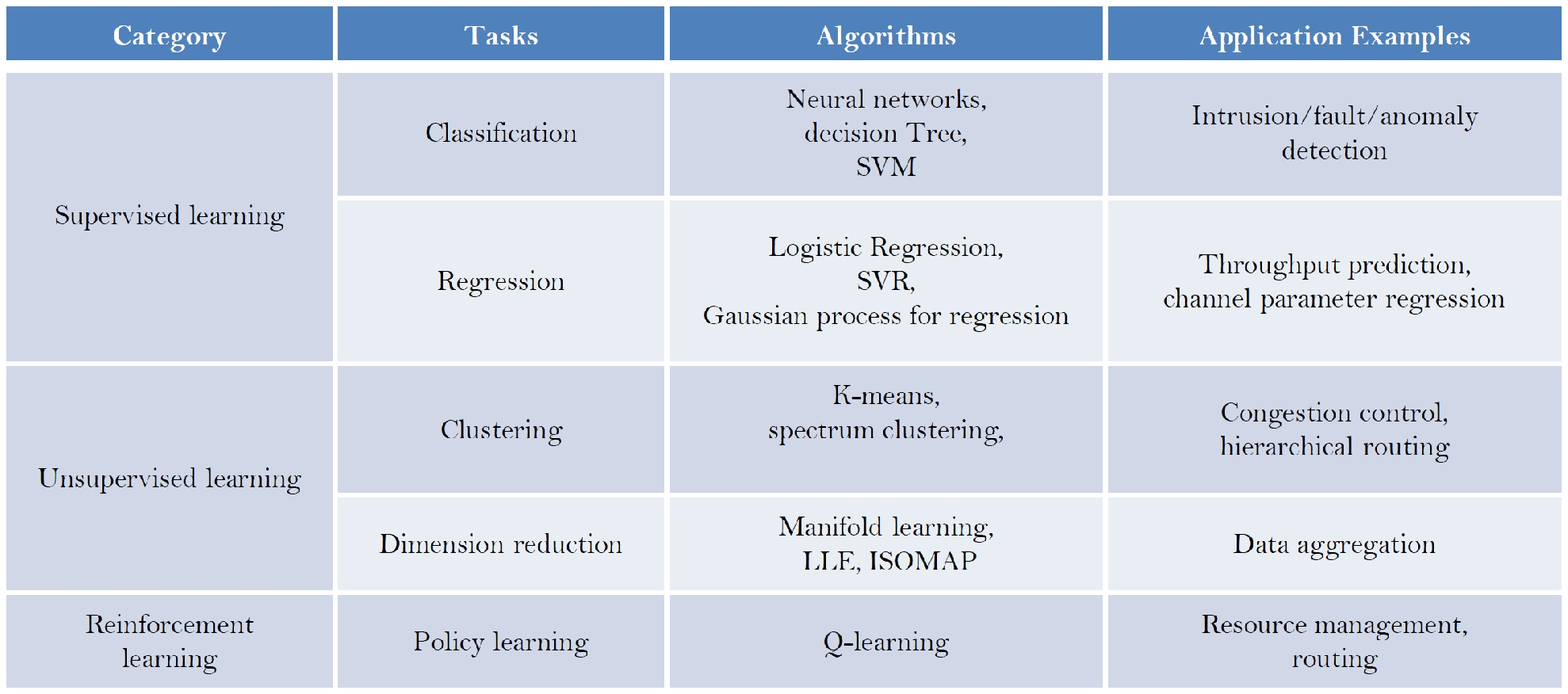}
\end{table}

\subsection{Supervised Learning}

Supervised learning learns from a labeled data set and can be further divided into classification and regression types.
Each training sample comes with a discrete (classification) or continuous (regression) value called a label or ground truth.
The ultimate goal of supervised learning is to learn the mapping from the input feature space to the label or decision space.

Classification algorithms assign a class label to each incoming sample.
In wireless networks, the classification problems include detecting whether the networks have been intruded or parts of the system are malfunctioning.
Some classic algorithms in this category include Bayesian classifiers, $k$-nearest neighbors (KNN), decision trees, support vector machine (SVM) and neural networks \cite{MLIntro}.

Instead of discrete outputs, regression algorithms predict a continuous value corresponding to each sample, such as estimating the house price given its associated feature inputs.
In wireless networks, the regression algorithms can potentially be used to predict channel parameters, network throughput, etc.
Classic algorithms include logistic regression, support vector regression (SVR), and Gaussian process for regression \cite{MLIntro}.

\subsection{Unsupervised Learning}

For supervised learning, with enough data, the error rate can be reduced close to the minimum error rate bound.
However, a large amount of labeled data is often hard to obtain in practice.
Therefore, learning with unlabeled data, known as unsupervised learning, has attracted more and more attention these days.
Unsupervised learning aims to find efficient representation of the data samples.
For instance, samples might be explained by hidden structures or hidden variables, which can be represented and learned by Bayesian learning methods.

Clustering is a representative problem of unsupervised learning, referring to grouping samples sharing similarities into different clusters.
Input features could be either the absolute description of each sample or the relative similarities between samples.
In wireless sensor networks, routing algorithms of hierarchical protocols need to cluster nearby nodes into a group since it is more energy-efficient for members within a cluster to send messages to a central node before inter-cluster transmission.
Classic clustering algorithms include $k$-means, hierarchical clustering, spectrum clustering, and Dirichlet process \cite{MLIntro}.

Another important class of unsupervised learning is dimension reduction, which projects samples from a high dimensional space into a lower one without losing too much information.
In many scenarios, the raw data come with high dimension and we may want to reduce the input dimension for several reasons. The first is due to the so called \emph{curse of dimensionality}, which describes the problematic phenomenon that arises when the dimension becomes huge. For example, in optimization, clustering, and classification, the model complexity and the number of required training samples dramatically grow with the feature dimension.
The second reason is that the inputs of each dimension are usually correlated and some dimensions may be corrupted with noise and interference, which will degrade the learning performance significantly if not handled properly.
A typical application example in wireless networks is the data aggregation performed by vehicular cluster heads before transmission to infrastructure nodes to reduce communication costs in cluster-based vehicular networks.
Some classic dimension reduction algorithms include linear projection methods, such as principal component analysis (PCA), and nonlinear projection methods, such as manifold learning, local linear embedding (LLE), and isometric mapping (ISOMAP) \cite{MLIntro}.

\subsection{Reinforcement Learning}
Reinforcement learning learns what to do, i.e., how to map situations to actions, through interacting with the environment in a trial-and-error search so as to maximize a reward. Therefore, it comes without explicit supervision.
Markov decision process (MDP) is generally assumed in reinforcement learning, which introduces actions and (delayed) rewards to the Markov process.
Learning $Q$ function is a classic model-free learning approach to solve the MDP problem, without the need for any information about the environment. This $Q$ function estimates the expectation of sum reward when taking an action in a given state, and the optimal $Q$ function is the maximum expected sum reward achievable by following any policy of choosing actions.
Reinforcement learning can be applied in vehicular networks to handle the temporal variation of wireless environments, which will be discussed in Section~\ref{sec:resource}.

\subsection{Deep Learning}

\begin{figure}[!t]
\centering
\includegraphics[width=0.7\linewidth]{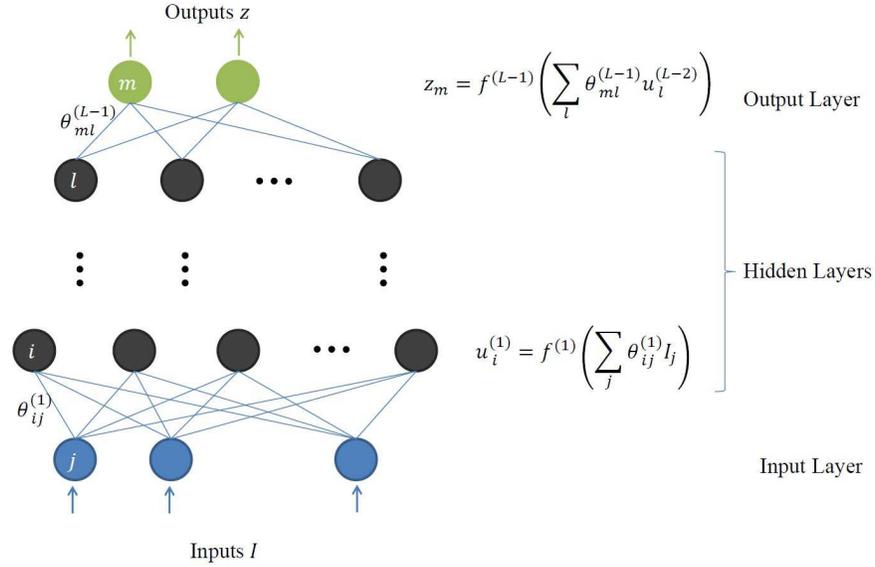}
\caption{An example of deep neural networks.}\label{fig:dl}
\end{figure}

Deep learning is a deeper version of neural networks, which consist multiple layers of neurons, as shown in Fig.~\ref{fig:dl}.
Recently, it has made significant advances on various machine learning tasks.
{Deep learning aims to learn the data representations, which can be built in supervised, unsupervised, and reinforcement learning.}
In Fig.~\ref{fig:dl}, the input layer is at the bottom, where each node in the figure represents a dimension of input data.
The output layer is at the top, corresponding to the desired outputs, whereas the layers in the middle are called hidden layers.
Each neuron in the network represents a non-linear transform, such as a sigmoid function or a Relu function, on a weighted sum of a subset of units in its lower layers.
Typically the number of hidden layers and the number of nodes in each layer are large since the neural network's representation ability grows as the hidden layers become deeper.
However, deeper networks bring new challenges, such as much more training data is needed and gradients of networks may easily explode or vanish \cite{DLIntro}.
With the help of faster computation resources, new training methods (new activation functions, pre-training), and new structures (batch norm, residual networks), training such deep architecture becomes possible.
Deep learning has been widely used in computer vision, speech recognition, natural language processing, etc, and greatly improved state-of-the-art performance in each area.
Depending on applications, different structures can be added to the deep networks.
For instance, convolutional networks share weights among spatial dimensions while recurrent neural networks (RNN) and long short term memory (LSTM) share weights among the temporal dimensions \cite{DLIntro}.

\section{Data-driven Decision Making in Vehicular Networks}
To overcome the unprecedented challenges encountered in vehicular networks, it is necessary to rethink traditional approaches to wireless network design, especially given the rich sources of data from various onboard sensors, road side monitoring facilities, historical transmissions, etc.
Indeed, it is highly desirable to devise efficient methods to interpret and mine the massive amounts of data and facilitate more data-driven decision making to improve vehicular network performance.
Machine learning represents an effective tool to serve such purposes with proven good performance in a wide variety of applications, as demonstrated by some preliminary examples in this section.

\subsection{Traffic Flow Prediction}
Timely and accurate acquisition of traffic flow information is a critical element of the ITS deployment since it lays out the foundation for many other services or applications, such as traffic congestion alleviation, fuel consumption reduction, and various location-based services.
The objective of traffic flow prediction is to infer from multiple sources of data, including historical and real-time traffic data, the traffic flow information for the wide variety of ITS-related applications.
Machine learning tools can be exploited to produce prediction outputs with accuracy that is hardly achievable using conventional approaches.
In \cite{ide2015lte}, a probabilistic graphical model, Poisson regression trees (PRT), has been used for two correlated tasks, the LTE communication connectivities prediction and the vehicular traffic prediction. The PRT is used for modeling the count data and is similar to decision trees, where each inner node represents the splitting criterion.
Information about the congestion and the performance of the communication system as well as the vehicular traffic information are used to enhance the prediction performance.
A novel deep-learning based traffic flow prediction method based on a stacked autoencoder model has further been proposed in \cite{lv2015traffic}, where autoencoders are used as building blocks to represent traffic flow features for prediction and achieve significant performance improvement.

\subsection{Local Data Storage in Vehicular Networks}
The position and the connectivity of vehicles are constantly changing in a vehicular environment.
However, some data, such as road status and camera sensor information, are region specific and can be used for local traffic information acquisition and estimation, which will be beneficial for load-balancing and user-behavior-based adjustment.
{In vehicular networks, data are naturally generated and stored across different units in the network, e.g., vehicles, road side units, remote clouds, etc. A framework has been developed in \cite{wu2017reinforcement} to store such data in vehicles without any support from infrastructure}. By using unicast transmission to transmit data between vehicles, the region specific data will always be kept in the region of interest. Initial selection of the next data carrier vehicle node is based on fuzzy logic instant evaluation, which is further refined through applying reinforcement learning.
As such, the data carrier node selection takes into account throughput, velocity, and bandwidth efficiency through the fuzzy logic-based short-term evaluation and also guarantees the long-term rewards through applying Q-learning. Reinforcement learning has been further applied to find efficient routing strategy to transfer data from the source node to the selected data carrier node.

\subsection{Network Congestion Control}
In the urban environment, the intersections are critical places where congestions of vehicles and the communication networks often take place.
A central controlled approach to manage congestion control at intersections has been presented by \cite{taherkhani2016centralized} with the help of a specific unsupervised learning algorithm, $k$-means clustering.
The approach basically addresses the congestion problem when vehicles stop at a red light in an intersection, where the road side infrastructures observe the wireless channels to measure and control channel congestion.
Transmission data are clustered into different groups through the use of a $k$-means clustering mechanism according to their features, including the message size, validity of messages, distances between vehicles and road side infrastructure, type of messages, and the direction of message senders.
Each cluster is provided with independent communication parameters, including transmission rate, transmission power, contention window size, i.e., the maximum backoff time, and Arbitration Inter-Frame Spacing (AIFS), i.e., the minimum time.
The channel has to be unoccupied before transmission so that collision is avoided.

\section{Intelligent Wireless Resource Management}\label{sec:resource}

Judicious management of various resources, such as spectrum, transmission power, storage, and computing, is critical for the proper functioning of vehicular networks.
Currently, a mainstream approach to resource management is to formulate an optimization problem and then obtain an optimal or sub-optimal solution depending on the performance-complexity tradeoff.
However, in practice, vehicular networks are highly dynamic, where the channels and network topology are constantly changing, leading to time-varying optimal solutions.
As a result, the optimization problem needs to be recomputed every time a small change occurs in the system, thus incurring huge network overhead.
Fortunately, reinforcement learning can serve as an effective alternative solution to the challenge, which learns to interact with the unknown environment, adapts to environmental changes, and takes proper actions. In this section, we will take a closer look at this topic.

\subsection{Load Balancing and Vertical Control}
There exist potential patterns and regularities of spatial-temporal distribution in traffic flow every day.
Reinforcement learning provides an effective tool to utilize such information to address the user association problem in a dynamic vehicular environment.
An online reinforcement learning approach for user association with load-balancing has been proposed in \cite{li2017user}, where an initial association decision is first made using reinforcement learning simply based on the current information.
Afterwards, the base station keeps accumulating such association information and uses historical association patterns to update the association results directly and adaptively.
Along this line, a fuzzy Q-learning based vertical handoff strategy for heterogeneous vehicular networks has been proposed in \cite{xu2014fuzzy}, which determines network connectivity based on four input parameters: received signal strength value, vehicle speed, data quantity, and the number of users associated with the targeted network.
The proposed learning-based strategy ensures seamless mobility management without the need for prior knowledge on handoff behavior.

\subsection{Virtual Resource Allocation}

\begin{figure}[!t]
\centering
\includegraphics[width=\linewidth]{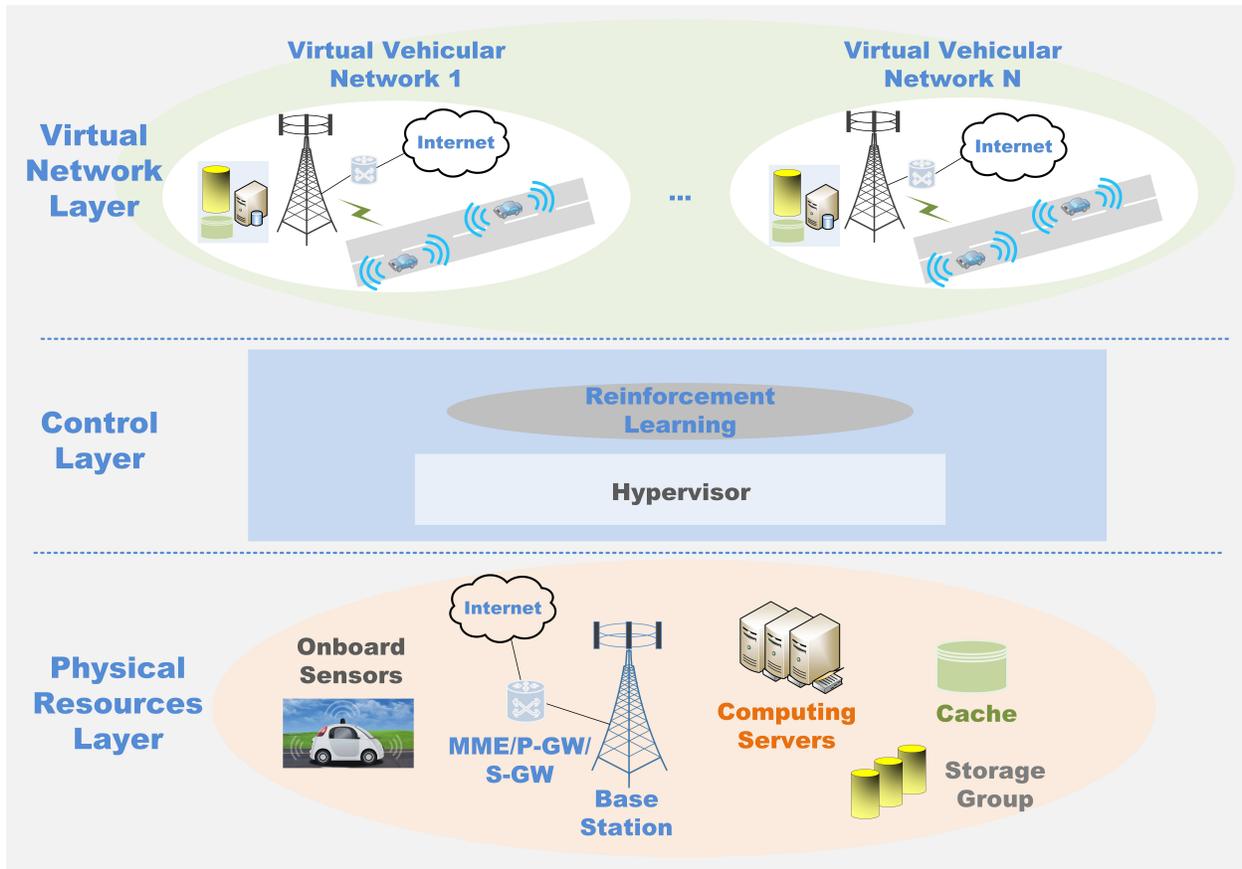}
\caption{An illustration of virtual vehicular networks.} \label{fig:virtual}
\end{figure}

Vehicular clouds, as shown in Fig.~\ref{fig:virtual}, consisting of various onboard units, road side units, and remote cloud servers, provide a pool of processing, sensing, storage, and communication resources that are dynamically provisioned for the ITS services.
How to dynamically allocate the resources to maximize the QoS for end users with small provisioning overhead is a nontrivial task.
In \cite{salahuddin2016reinforcement}, the resource allocation problem is modeled as an MDP, where the set of states contains all possible configurations of the allocated resources and the actions are defined as the transition from one state to another.
A reinforcement learning framework has been proposed for resource provisioning to cater for such dynamic demands of resources while respecting stringent QoS requirements.
A two-stage delay optimal dynamic virtualization radio scheduling scheme has been developed in \cite{zheng2016delay}, which takes a joint consideration of the large timescale factors, such as the traffic density, and short-term factors, such as the channel and queue state information.
The dynamic delay-optimal virtualization radio resource management has been formulated as an partially observed MDP, which is then solved through an online distributed learning approach.
A unified framework has been proposed in \cite{he2017integrated} for dynamic orchestration of networking, caching and computing. The resource allocation problem in the unified framework is formulated as a joint optimization problem. To deal with the high complexity of the joint optimization problem, a deep reinforcement learning approach has been proposed and desirable performance has been demonstrated.
{In the future, network slicing will be built on the virtual networks so that the logical network functions and the parameter configurations can be tailored to meet the requirement of a specific service. Learning the slicing management according to the arrival traffic will be essential to support different use cases in the vehicular networks.}

\subsection{Distributed Resource Management}
\begin{figure}[!t]
\centering
\includegraphics[width = 0.9 \linewidth]{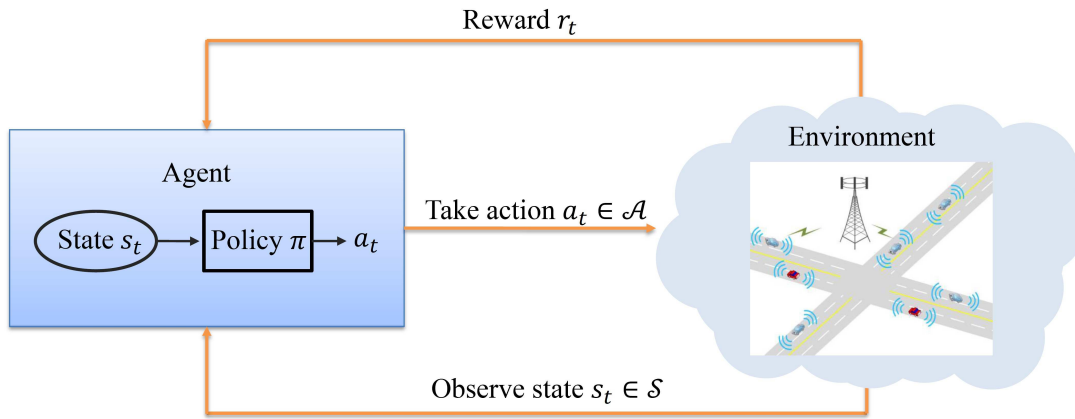}
\caption{Deep reinforcement learning for V2V communications.} \label{fig:rl}
\end{figure}

There have been many interesting works on resource allocation for device-to-device (D2D) based vehicular communications.  Most of them are centralized, where the central controller collects information and makes decisions for all the vehicles.
Nevertheless, centralized control schemes will incur huge overheads in order to acquire the global network information, which grows dramatically with the number of vehicular links.
As shown in Fig.~\ref{fig:rl}, we have developed a decentralized resource allocation mechanism for vehicular networks based on deep reinforcement learning \cite{Ye2017re}, which is used to find the mapping between the partial observations of each vehicle agent and the optimal resource allocation solution.
In particular, this method can address stringent latency requirement on vehicle-to-vehicle (V2V) links, which is usually hard to deal with using existing optimization approaches.

We assume the vehicle-to-infrastructure (V2I) link has been allocated orthogonal resources beforehand and the main goal of the proposed distributed channel and power allocation is to ensure the latency constraints for each V2V link and minimize interference to V2I links.
The structure of reinforcement learning for V2V communications is shown in Fig.~\ref{fig:rl}, where an agent, corresponding to a V2V link, interacts with the environment.
In this scenario, the environment is considered to be everything outside the V2V link.
It should be noted that the behaviors of other V2V links cannot be controlled in the decentralized settings.
As a result, their actions, such as selected spectrum, transmission power, etc., can only be treated as part of the environment.

As in Fig.~\ref{fig:rl}, at time $t$, each V2V link, as an agent, observes a state, $s_t$, from the state space, $\mathcal{S}$, and accordingly takes an action, $a_t$, selected from the action space, $\mathcal{A}$,  which amounts to selecting the sub-band and transmission power based on the policy, $\pi$.
The decision policy, $\pi$, is determined by a Q-function, $Q(s_t, a_t, \theta)$, where $\theta$ is the parameter of the Q-function and can be obtained by deep learning.
Following the action, the state of the environment transitions to a new state, $s_{t+1}$, and the agent receives a reward, $r_t$, determined by the capacity of the V2I link and the corresponding latency.
In our system, the state observed by each V2V link for characterizing the environment consists of several parts: the instantaneous channel information of the corresponding V2V link, $g_t$, the previous interference to the link, $I_{t-1}$, the channel information of the V2I link, i.e., from the V2V transmitter to the base station, $h_t$, the selection of sub-bands of neighbors in the previous time slot, $B_{t-1}$, the remaining load for the vehicles to transmit, $L_t$, and the remaining time to meet the latency constraints $U_t$. Hence the state can be expressed as $s_t = [g_t, I_{t-1}, h_t, B_{t-1}, L_t, U_t]$.
The training and testing data are generated from an environment simulator based on 3GPP channel models.
In the training stage, we follow the deep Q-learning with experience replay, where the generated data are saved in a storage called \emph{memory}. The mini-batch data used for updating the Q-network is sampled from the \emph{memory} in each iteration.
In this way, the temporal correlation of data can be suppressed.
The policy used in each V2V link for selecting spectrum and power is random at the beginning and gradually improved with the updated Q-networks.

\begin{figure}[!t]
\centering
\includegraphics[width=.7\linewidth]{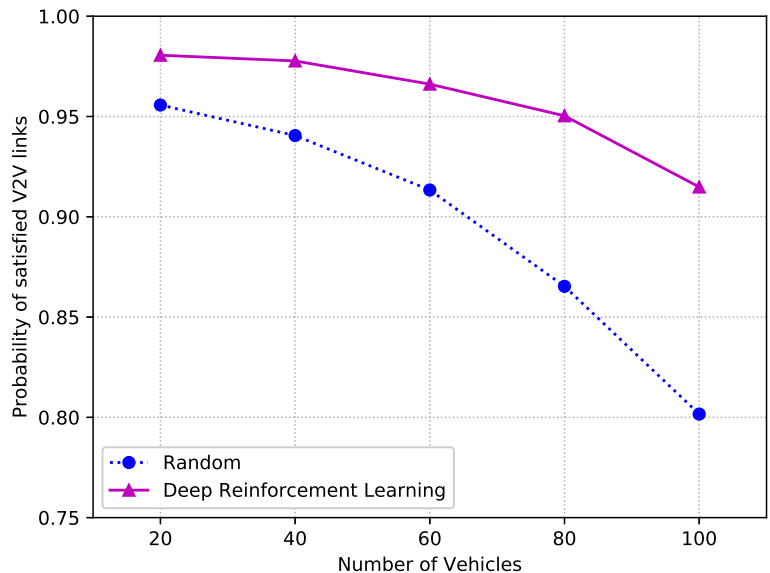}
\caption{Success probability versus the number of vehicles.} \label{fig:V2V}
\end{figure}

Fig.~\ref{fig:V2V} compares the proposed method with a random resource allocation method, where the agent randomly chooses a sub-band for transmission at each time.
From the figure, the above reinforcement learning based method has a larger probability for V2V links to satisfy the latency constraint since it can dynamically adjust the power and sub-band for transmission so that the links that are likely to violate the latency constraint have more resources.


\section{Open Issues}

Recent hyper around machine learning seems suggesting it is a panacea to most conventionally challenging problems, especially in view of the significant advances made by deep learning. However, naively applying existing machine learning methods to vehicular networks is expected to be insufficient due to their many distinguishing characteristics.
How to adapt existing learning methods or develop V2X specific learning algorithms to better handle such characteristics remains a challenging task.
Therefore, in this section, we identify several research topics for further investigation.

\subsection{Learning Dynamics of Vehicular Networks}
Vehicular networks exhibit strong dynamics in many facets, e.g., wireless propagation channels, network topologies, traffic dynamics, etc.
How to efficiently learn and robustly predict such dynamics based on historical data generated from multiple onboard sensors or previous transmission is still an open issue.
{Traditionally, some Bayesian models, such as the hidden Markov models, can be used for characterizing the temporal relationship and predicting states in the next time slot.
New sophisticated models powered by deep neural networks, such as RNN and LSTM, can improve the prediction by exploiting the long-range dependency.
One potential application is to predict wireless channels using deep neural network based on the received signal and historical data. Due to the high mobility of users and advanced techniques being adopted, such as massive multiple-input multiple-output (MIMO) and millimeter wave communications, new challenges arise on the task of estimating the high dimensional fast-varying wireless channel.
Deep learning has shown strong abilities to efficiently distill high dimensional data by exploiting such properties as sparsity. It is unclear whether deep neural network can assist or even replace the existing channel estimation module that requires transmitting frequent pilot symbols to track channel variations.}
{Another potential application is to predict the vehicular trajectory, which could be further used for traffic dynamics prediction. The latent factors that affect the trajectories, such as driver’s intention, traffic situations, and road structure, may be implicitly learned from the historical data based on deep neural networks.}
More research efforts are thus needed to develop a better understanding in this area.

\subsection{Method Complexity}
Conventional machine learning techniques require much work on designing feature representation while deep learning provides an effective way to learn features from raw data.
With deep models, information can be distilled more efficiently than the traditional methods and experimental results have confirmed that the deep hierarchical structure is necessary for better understanding of data.
Key technologies have since been devised to get deeper models with stronger representation abilities.
With high performance computing facilities, such as graphics processing unit (GPU), deeper networks can be trained with massive amounts of data through advanced training techniques, such as batch norm and residual networks \cite{DLIntro}.
However, in vehicular networks, onboard computation resources are limited and the low end-to-end latency requirement restricts heavy use of cloud servers housed remotely.
It is therefore important to develop special treatments for vehicular networks, such as model reduction or compression, to relieve the resource limitation without compromising performance.

\subsection{Distributed Representation}
In vehicular networks, data are naturally generated and stored across different units in the network, e.g., vehicles, road side units, remote clouds, etc.
This brings challenges to the applicability of most existing machine learning algorithms that have been developed under the assumption that data are centralized controlled and easily accessible.
As a result, distributed learning methods are desired in vehicular networks that act on partially observed data and have the ability to exploit information obtained from other entities in the network.
Furthermore, additional overheads incurred by the coordination and sharing of information among various units in vehicular networks for distributed learning shall be properly accounted for to make the system work effectively.

\section{Conclusion}
This article provides an overview of applying machine learning to address challenges in emerging vehicular networks.
We have briefly introduced the basics of machine learning, including major categories and representative algorithms.
We have then provided some preliminary examples in applying machine learning in vehicular networks to facilitate data-driven decision making, and discussed intelligent wireless resource management using reinforcement learning in detail.
We have finally highlighted some open issues for further research.

\section*{Acknowledgement}
This work was supported in part by a research gift from Intel Corporation and the National Science Foundation under Grants 1443894 and 1731017.

\bibliographystyle{IEEEtran}
\bibliography{ml_v2x}

\end{document}